\documentclass[conference,letterpaper,twoside,final,10pt]{ieeeconf}

\usepackage{graphicx}
\usepackage{amssymb}
\usepackage{amsmath}
\usepackage{cite}
\usepackage{comment}
\usepackage{color}
\usepackage{url}
\usepackage{alltt}
\usepackage{cleveref}
\usepackage{tikz}
\usepackage{pgfplots}
\usepackage{pgf-umlsd}
\usepackage{multirow}
\usepackage{algorithm}
\usepackage[noend]{algpseudocode}
\usepackage{bigfoot}
\usepackage{paralist}
\usepackage{tabularx}
\usepackage[whole]{bxcjkjatype}
\usepackage{ifthen}
\usepackage{threeparttable}
\usepackage[hang,small,bf]{caption}
\usepackage[subrefformat=parens]{subcaption}
\pgfplotsset{compat=1.14}
\IEEEoverridecommandlockouts
\DeclareMathOperator*{\argmin}{argmin}

\newcommand{\argmax}{\mathop{\rm arg~max}\limits}

\title{\LARGE \bf
  DQN-TAMER: Human-in-the-Loop Reinforcement Learning with Intractable Feedback
}
\author{
Riku Arakawa$^{*}$$^{\dagger}$,
Sosuke Kobayashi $^{\dagger}$,
Yuya Unno $^{\dagger}$,
Yuta Tsuboi $^{\dagger}$,
Shin-ichi Maeda $^{\dagger}$
\thanks{$^{*}$ The University of Tokyo. \newline \url{arakawa-riku428@g.ecc.u-tokyo.ac.jp}}
\thanks{$^{\dagger}$ Preferred Networks, Inc. \newline \url{{sosk, unno, tsuboi, ichi}@preferred.jp}}}

\begin{document}
\maketitle

\begin{abstract}

Exploration has been one of the greatest challenges in reinforcement learning (RL), which is a large obstacle in the application of RL to robotics.
Even with state-of-the-art RL algorithms, building a well-learned agent often requires too many trials, mainly due to the difficulty of matching its actions with rewards in the distant future.
A remedy for this is to train an agent with real-time feedback from a human observer who immediately gives rewards for some actions.
This study tackles a series of challenges for introducing such a human-in-the-loop RL scheme.
The first contribution of this work is our experiments with a precisely modeled human observer: \textsc{binary}, \textsc{delay}, \textsc{stochasticity}, \textsc{unsustainability}, and \textsc{natural reaction}.
We also propose an RL method called DQN-TAMER, which efficiently uses both human feedback and distant rewards.
We find that DQN-TAMER agents outperform their baselines in Maze and Taxi simulated environments.
Furthermore, we demonstrate a real-world human-in-the-loop RL application where a camera automatically recognizes a user's facial expressions as feedback to the agent while the agent explores a maze.

\end{abstract}

\section{Introduction}

Reinforcement learning (RL) has potential applications for autonomous robots~\cite{DBLP:journals/ijrr/KoberBP13}.
Even against highly complex tasks like visuomotor-based manipulation~\cite{DBLP:journals/jmlr/LevineFDA16} and opening a door with an arm~\cite{DBLP:conf/icra/GuHLL17}, skillful policies for robots can be obtained through repeated trials of deep RL algorithms.

However, exploration remains as one of the greatest challenges, preventing RL from spreading to real applications.
It often requires a lot of trials until the agent reaches an optimal policy.
This is primarily because RL agents obtain rewards only in the distant future, e.g., at the end of the task. Thus, it is difficult to propagate the reward back to actions that play a vital part in receiving the reward.
The estimated values of actions in given states are modified exponentially slowly over the number of remaining intervals until the future reward is received~\cite{DBLP:journals/corr/abs-1806-07857}.

Additional training signals from a human are a very useful remedy.
One direction involves human demonstrations.
Using human demonstrations for imitation learning can efficiently train a robot agent~\cite{DBLP:journals/corr/abs-1709-10089},
though it is sometimes difficult or time-consuming to collect human demonstrations.

We use real-time feedback from human observers as another helpful direction in this study.
During training, human observers perceive the agent's actions and states in the environment and provide some feedback to the agent in real time rather than at the end of each episode.
Such immediate rewards can accelerate learning and reduce the number of required trials.
This method is called \textit{human-in-the-loop} RL and its effectiveness has been reported in prior publications~\cite{Thomaz2005RealTimeIR,DBLP:conf/ro-man/ThomazHB06,DBLP:conf/aihc/Broekens07,4640845,DBLP:conf/iberamia/Tenorio-GonzalezMP10,5975338,NIPS2013_5187,DBLP:conf/icml/MacGlashanHLPWR17,Arumugam2018DeepRL,DBLP:conf/aaai/WarnellWLS18 }.

\begin{figure}[t]
 \begin{center}
	\includegraphics[width=80mm]{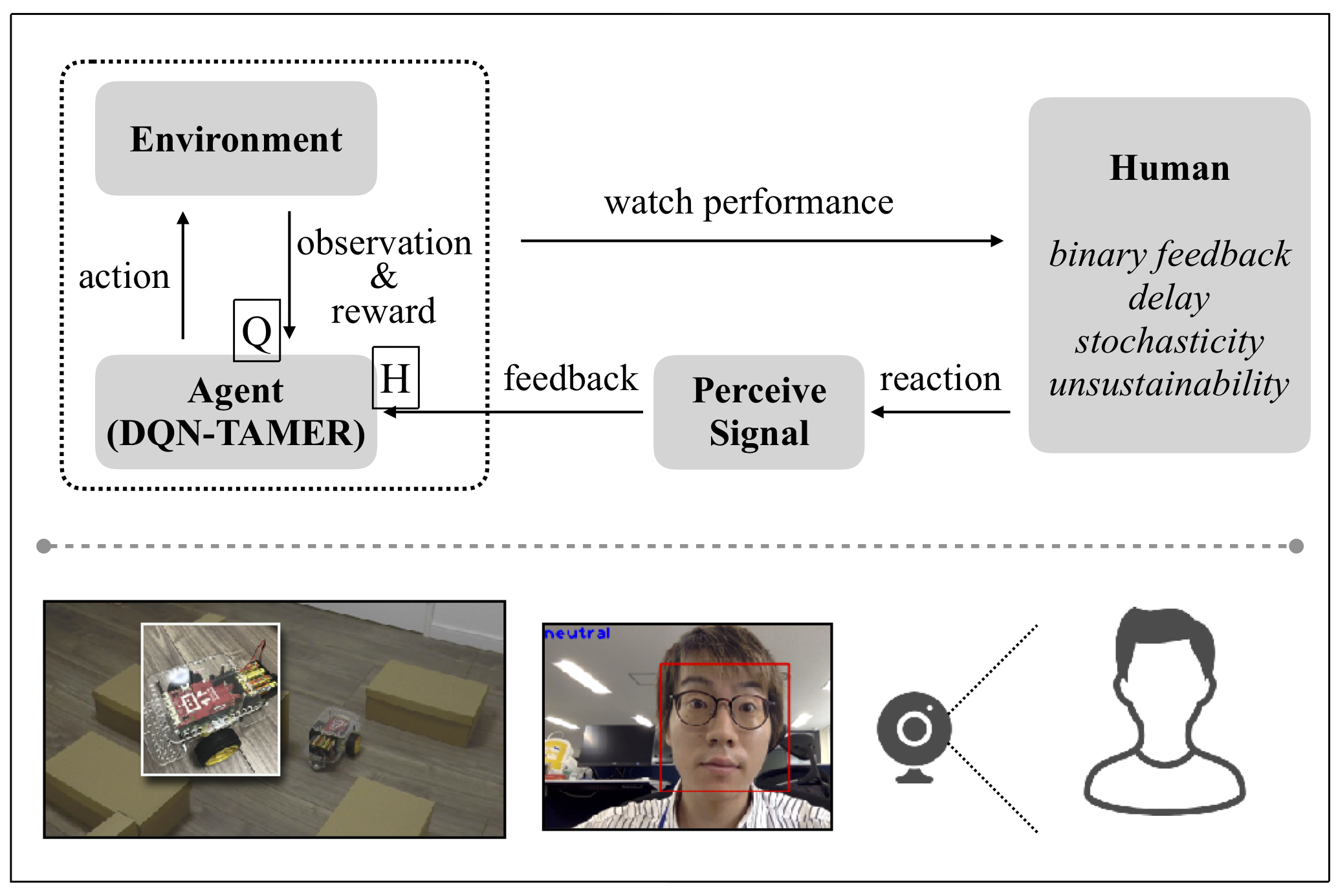}
 \end{center}
 \caption{Overview of human-in-the-loop RL and our model (DQN-TAMER). The agent asynchronously interacts with a human observer in the given environment. DQN-TAMER decides actions based on two models. One (Q) estimates rewards from the environment and the other (H) for feedback from the human.}
 \label{fig:overview}
\end{figure}

Human-in-the-loop RL has the potential to greatly improve training thanks to the immediate rewards.
However, experiments in prior studies did not consider some key factors in realistic human-robot interactions.
They sometimes assumed that human observers could (1) give precise numerical rewards, (2) do so without delay (3) at every time step, and (4) that rewards would continue forever.
In this paper, we reformulate human observers with the following more realistic characteristics: binary feedback, delay, stochasticity, and unsustainability.
Furthermore, we examine the effect from recognition errors,
when an agent autonomously infers implicit human reward from natural reactions like facial expressions.
Table~\ref{tab:prior_works} shows a comparison with prior work.

\begin{table*}[t]
  \centering
  \caption{Characteristics of human observers tested in prior work and this study
}

  \begin{tabular}{cccccc}\hline
	study & \textsc{binary} & \textsc{delay} & \textsc{stochasticity} & \textsc{unsustainability} & \textsc{natural reaction} \\ \hline
   Andrea et al. 2005\cite{Thomaz2005RealTimeIR,DBLP:conf/ro-man/ThomazHB06} && $\checkmark$ & $\checkmark$ &&\\
   Joost Broekens 2007\cite{DBLP:conf/aihc/Broekens07} 
 & $\checkmark$ &&&$\checkmark$ & $\checkmark$ {\rm (facial expression)} \\
 	Knox et al. 2007 \cite{4640845}
 & $\checkmark$ & $\checkmark$ & $\checkmark$ && \\
   Tenorio-Gonzalez et al. 2010 \cite{DBLP:conf/iberamia/Tenorio-GonzalezMP10}
 & & $\checkmark$ & $\checkmark$ && $\checkmark$ {\rm (voice)} \\
  	Pilarski et al. 2011\cite{5975338} 
 & $\checkmark$ & $\checkmark$ & $\checkmark$ && \\
 	Griffith et al. 2013 \cite{NIPS2013_5187} 
 & $\checkmark$ && $\checkmark$ && \\
   MacGlashan et al. 2017 \cite{DBLP:conf/icml/MacGlashanHLPWR17}
 && $\checkmark$ & $\checkmark$ &$\checkmark$ & \\
 	Arumugam et al. 2018\cite{Arumugam2018DeepRL} 
 && $\checkmark$ & $\checkmark$ &$\checkmark$ & \\
 	Warnel et al. 2018 \cite{DBLP:conf/aaai/WarnellWLS18}
 & $\checkmark$ & $\checkmark$ & $\checkmark$ && \\  \hline
 	 {\bf Ours} 
 & $\checkmark$ & $\checkmark$ & $\checkmark$ &$\checkmark$ & $\checkmark$ {\rm (facial expression)} \\
  \end{tabular}
  \label{tab:prior_works}
\end{table*}

With such a human-in-the-loop setup,
we derive an efficient RL algorithm called DQN-TAMER from an existing human-in-the-loop algorithm (TAMER) and deep Q-learning (DQN).
The DQN-TAMER algorithm learns two disentangled value functions for the human immediate reward and distant long-term reward.
DQN-TAMER can be seen as a generalization of TAMER and DQN, where the contribution from each model can arbitrarily controlled.

The contributions of the paper are as follows:
\begin{enumerate}

\item We precisely formulate the following more realistic human-in-the-loop RL settings: (\textsc{binary feedback}, \textsc{delay}, \textsc{stochasticity}, \textsc{unsustainablity}, \textsc{natural reaction}).

\item We propose an algorithm, DQN-TAMER, for human-in-the-loop RL, and demonstrate that it outperforms the existing RL methods in two tasks with a human observer.

\item We built a human-in-the-loop RL system with a camera, which autonomously recognized a human facial expression and exploited it for effective explorations and faster convergence.

\end{enumerate}

\section{Problem Formulation \label{sec: problem_formulation}}

We first describe the standard RL settings and subsequently introduce a human observer for human-in-the-loop RL, as shown in Figure~\ref{fig:overview}.
We then describe the characteristics of the human observer.

In standard RL settings, an agent interacts with an environment $\mathcal{E}$ through a sequence of observations, actions, and rewards.
At each time $t$, the agent receives an observation $s_t$ from $\mathcal{E}$, takes an action $a_t$ from a set of possible actions $\mathcal{A}$, and then obtains a reward $r_{t+1}$.
Let $\pi(a|s)$ be the trainable policy of the agent for choosing an action $a$ from $\mathcal{A}$ given an observed state $s$.
The ultimate goal of RL is to find the optimal policy which maximizes the expectation of the total reward $R_t = \sum_{k=0}^{\infty} \gamma^k r_{t+k}$ at each state $s_t$, where $\gamma$ is a discount factor for later rewards.

Next, we consider introducing a human into the above RL settings. At each step, a human watches the agent's action $a_t$ and the next state $s_{t+1}$, assesses $a_t$ based on intuition or some other criteria, and gives some feedback $f_{t+1}$ to the agent through some type of reaction.
Prior work has explored modeling human feedback.
This study discusses and reformulates those clearly as five components.
This paper is the first study that fully integrates all components and performs experiments and analysis to test their effects.

\subsection{Binary}
Some studies consider humans giving various values as feedback to influence the agent~\cite{Thomaz2005RealTimeIR,DBLP:conf/ro-man/ThomazHB06}.
However, requesting people give fine-grained or continuous scores is found difficult~\cite{PRESTON20001} 
because it requires human have enough understanding of the task at hand and requires that human can rate the agent behavior quantitatively in an objective manner.
This is why binary feedback is preferred.
The feedback simply indicates whether an action is good or bad.
In this way, even an ordinary person can be a desirable observer and provide feedback as well as an expert~\cite{DBLP:conf/nips/ChristianoLBMLA17}.
Thus, we assume binary feedback, i.e., $f_t \in \{-1, +1\}$.

\subsection{Delay}
One may think that human feedback will surely accelerate an agent's learning.
In realistic settings, however, it is actually difficult to utilize feedback because human feedback is usually delayed by a significant amount of time~\cite{hockley1984analysis}.
In particular, the agent must perform actions in a dynamic environment where the state changes continuously. Thus, the agent cannot wait for feedback at each step.
Furthermore, the delay must not be constant,
implicitly depending on people's concentration, complexity of states, and actions, etc.
The randomness of delay makes the problem much more difficult.
We assume that the number of feedback delay steps follows a certain probability distribution.
%
%
%

\if0
We introduce $p_{\rm delay}(i)$, the probability of delaying in $i$ steps, such that 
$p_{\rm delay}(i) = 1/(n+1)$ for $i\in\{0,\cdots,n\}$ and $p_{\rm delay}(i) = 0$ for $i > n$ where $n$ denotes the maximum time step that the delay could happen.
\fi
\if0
$\sum_{i=0}^{\infty} p_{\rm delay}(i) = 1$.
It is reasonable to suppose this distribution is similar to that reported by Hockley~\cite{hockley1984analysis}.
In detail, we derive $p_{\rm delay}(i)$ as follows:
\begin{eqnarray}
p_{\rm delay}(i) &=& P( i\times d \leq t < (i+1)\times d) \nonumber \\
&=& \int_{i\times d }^{(i+1)\times d} f(\tau) d\tau,
\end{eqnarray}
where $f(\tau)$ is the probability density function based on Hockley's result and $d$ is the time interval between two actions of the agent.
In the following experiments, we fix that $p_{delay}(0) = 0.3$, $p_{delay}(1) = 0.6$, $p_{delay}(2) = 0.1$, $p_{delay}(n) = 0 \,\,(n \geq 3)$.
\fi
%

Surprisingly, we found that human feedback could have totally ``negative'' effects on the existing learning algorithms of an agent if the agent ignores this delay effect and takes the feedback as exact and immediate feedback.
On the other hand, our proposed learning algorithm succeeds when such delayed human feedback is utilized
even though the actual probability of delay is different from the one we assumed.

\subsection{Stochasticity}

In addition to delaying feedback, other studies missed the idea that people could not always give feedback when an agent performs an action correctly.
It is also reported that the feedback frequency varies largely among human users~\cite{DBLP:conf/nips/IsbellSKSS01,DBLP:conf/agents/IsbellSKSS01}.
Thus, such a stochastic drop is a factor of intractable human feedback that we have to model for human-in-the-loop RL.

We introduce $p_{\rm feedback}$ to indicate the probability that appropriate feedback occurs in a time step (i.e. the probability of avoiding drop) to model the difficulty of random events.
We vary the strength of stochasticity in the following experiments and confirm a significant effect in learning process.

\subsection{Unsustainability}
Even after introducing delay and stochasticity, the setting is still less realistic.
It is very difficult to presume that humans watch an agent until it finishes learning through many episodes.
The learning process might last a long time, thus a human may leave before the agent converges to an optimal policy.
Ideally, even if a human gives feedback within a limited span after learning begins, we wish it could subsequently lead to a better learning process.
Here we introduce the notion of feedback stop with a time step $t_{\rm stop}$, where the human leaves the environment and the agent stops receiving feedback.
We confirm that ending feedback degrades learning process of prior algorithms; in contrast, our proposed algorithm works robustly.

\subsection{Natural Reaction}
Finally, the method used to provide feedback is not unique or obvious.
One naive method for providing binary feedback is using positive-negative buttons or levers.
However, when intelligent agents become more ubiquitous and we launch real human robot interaction systems, it is preferable that the system infer implicit feedback from natural human reactions rather than humans actively providing feedback.
Robots with such a mechanism would be capable of lifelong learning~\cite{Thrun1995LifelongRL,finn2016generalizing} after deployment in the real world.
For example, robot pets might utilize their owner's voice as feedback for directions or some toy tasks, or communication robots possibly infer feedback from a user via their facial expressions.

In this paper, we investigated the use of human facial expressions.
We use a deep neural network-based classifier for facial expression recognition and we built a demo system with a camera.
Note that classification errors from such a model cause an agent to misunderstand the sentiment polarity (positive or negative) associated with feedback.
This is another important issue which we believe will arise in future human robot interaction applications.

\section{Methods \label{sec:method}}
We first describe two existing RL algorithms.
Each algorithm is a well-known deep RL method.
We then propose an algorithm that generalizes both of them.

\subsection{Deep Q-Network (DQN)}

The optimal policy can be characterized as the policy that causes the agent to take and action that maximizes the action value for the action in the given state ~\cite{Watkins1992, Sutton1988}.
An action value function $Q_{\pi}: \mathcal{S} \times \mathcal{A} \rightarrow \mathcal{R}$ is a function that returns the expected total reward in a given state and for a given action when following the policy $\pi$~\cite{DBLP:books/lib/RussellN03}.
The optimal action value is defined as the maximum action value function with respect to the policy.
\begin{equation}
{Q}^\star(s, a) = \max_{\pi} {Q}_{\pi}(s, a).
\end{equation}
Q-learning is an algorithm that estimates the optimal action value
by iteratively updating the action value function using the Bellman update \cite{Watkins1992}.

A deep Q-network (DQN) is a kind of approximate Q-learning that
utilizes a deep neural network to represent the action value function 
together with some tricks in training, such as experience replay~\cite{Lin1992}, reward clipping, and a target network for stabilizing training~\cite{DBLP:journals/corr/MnihKSGAWR13}.

To handle the human feedback in the framework of RL, 
we augment an extra reward function that computes a scalar reward for human feedback in addition to the original reward function, i.e., we employ so-called \textit{reward shaping} \cite{DBLP:conf/icml/NgHR99,DBLP:conf/iui/KnoxS13} to incorporate human feedback.

\subsection{Deep TAMER}

TAMER \cite{4640845} is a current standard framework in human-in-the-loop RL, where the agent predicts human feedback and takes the action that is most likely to result in good feedback.
In short, TAMER is a value-based RL algorithm where the values are estimated from human feedback only.
Deep TAMER \cite{DBLP:conf/aaai/WarnellWLS18} is an algorithm that applies a deep neural network within this TAMER framework.

In Deep TAMER, the H-function is used instead of the Q-function to show the value of an action at a certain state ($H: \mathcal{S} \times \mathcal{A} \rightarrow \mathcal{R}$).
The differences from Q-learning is that H-function estimates a binary human feedback $f$ for each action.
Similar to DQN, and given the current estimate $\hat{H}$, the agent policy is
\begin{equation}
\pi(s)_{\rm Deep TAMER} = \argmax_a \: \hat{H}(s, a).
\label{eq:h_policy}
\end{equation}

Deep TAMER considers a certain feedback that corresponds to some recent state and action pairs, which expects \textsc{delay}.
Let $\boldsymbol{s}$ and $\boldsymbol{a}$ be a sequence of states and actions, respectively.
The loss function $L$ for judging the quality of $\hat{H}$ is defined as follows:
\begin{equation}
 L(\hat{H}; \boldsymbol{s}, \boldsymbol{a}, f) = \sum_{s \in \boldsymbol{s}, a \in \boldsymbol{a}} || \hat{H}(s, a) - f ||^2,
\end{equation}
The optimal feedback estimation is the value of $\hat{H}$ that minimizes the expected loss value, and Deep TAMER updates this using stochastic gradient descent (SGD):
\begin{equation}
\hat{H}^\star_{\pi}(s, a) = \argmin_{\hat{H}} \mathbb{E}_{\boldsymbol{s}, \boldsymbol{a}}[L(\hat{H}; \boldsymbol{s}, \boldsymbol{a}, f)]
\end{equation}
\begin{equation}
	\hat{H}(s, a)_{k+1} = \hat{H}(s, a)_{k} - \eta_k \nabla_{\hat{H}} L(\hat{H}_k; \boldsymbol{s}, \boldsymbol{a}, f)
\label{eq:h_update}
\end{equation}
where $\eta_k$ is the learning rate at update iteration $k$.

Also, inspired by experience replay in DQN \cite{Lin1992}, a similar technique is introduced to stabilize learning by the $\hat{H}$ neural network.
$D_{\rm local}$ is a set of tuples for a state, action, and feedback when a single feedback $f$ is received, which is defined as
\begin{equation}
D_{\rm local} = \{(s, a, f) \| (s, a) \in (\boldsymbol{s}, \boldsymbol{a})\}.
\label{eq:d_local}
\end{equation}
$D_{\rm global}$ stores all the past states, actions, and feedback pairs. Every time a new feedback occurs, it updates as follows:
\begin{equation}
D_{\rm global} \leftarrow D_{\rm global} \cup D_{\rm local}
\label{eq:d_global}
\end{equation}

\begin{algorithm}[t]
\caption{Deep TAMER}
\begin{algorithmic}
\Require initialized $\hat{H}$, update interval $b$, learning rate $\eta$
\Ensure $D_{\rm global}=\emptyset$
\While{NOT goal or time over} 
	\State observe $s$ 
	\State execute $a \sim \pi(s)_{\rm Deep TAMER} $ by (\ref{eq:h_policy})
	\If{new feedback $f$}
		\State prepare $ \boldsymbol{s}, \boldsymbol{a}$
		\State obtain $D_{\rm local}$ by (\ref{eq:d_local})
		\State update $D_{\rm global}$  by (\ref{eq:d_global})
      \State update $\hat{H}(s, a)$ by (\ref{eq:h_update}) using $D_{\rm local}$
	\EndIf
	\If{every $b$ steps and $D_{\rm global} \neq \emptyset$}
		\State update $\hat{H}(s, a)$ by (\ref{eq:h_update}) using mini-batch sampling from $D_{\rm global}$
	\EndIf
\EndWhile
\end{algorithmic}
\end{algorithm}

The TAMER framework (including Deep TAMER) only exploits human feedback and lacks the ability to make use of rewards from the environment.
Our proposed method is described in the next subsection, where the agent successfully uses both human feedback and environmental rewards.

\subsection{Proposed DQN-TAMER}

Our motivation lies in integrating the TAMER framework into an existing value based on Q-learning and, therefore, achieves faster agent learning convergence.

DQN-TAMER trains the Q-function and H-function separately using the DQN and Deep TAMER algorithms.
Given the estimated $\hat{Q}$ and $\hat{H}$ respectively, the agent policy is defined as
\begin{equation}
\pi(s)_{\rm DQN-TAMER} = \argmax_a \alpha_q \hat{Q}(s, a)+\alpha_h \hat{H}(s, a),
\label{eq:hq_policy}
\end{equation}
where $\alpha_q$ and $\alpha_h$ are the hyper parameters that determine the extent to which the agent relies on the reward from the environment and feedback from a human.
Note that, $\alpha_h$ decays at every step and eventually $\alpha_h \rightarrow 0$, thus the agent initially explores efficiently by following human feedback and eventually reaches the optimal DQN policy much faster.

\begin{algorithm}[t]
\caption{Proposed: DQN-TAMER}         
\begin{algorithmic}                  
\Require initialized $\hat{H}$, $\hat{Q}$, update interval $b$, learning rate $\eta$, weight $\alpha_q, \alpha_h$
\Ensure $D_{\rm global}=\emptyset$
\While{NOT goal or time over} 
	\State observe $s$ 
	\State execute $a \sim \pi(s)_{\rm DQN-TAMER} $ by (\ref{eq:hq_policy})
   \State decay $\alpha_h$
	\If{new feedback $f$}
		\State prepare $ \boldsymbol{s}, \boldsymbol{a}$
		\State obtain $D_{\rm local}$ by (\ref{eq:d_local})
		\State update $D_{\rm global}$  by (\ref{eq:d_global})
      \State update $\hat{H}(s, a)$ by (\ref{eq:h_update}) using $D_{\rm local}$
	\EndIf
	\If{every $b$ steps}
      \State update $\hat{Q}(s, a)$
      \If{$D_{\rm global} \neq \emptyset$}
      		\State update $\hat{H}(s, a)$ by (\ref{eq:h_update}) using mini-batch sampling from $D_{\rm global}$
      \EndIf
	\EndIf
\EndWhile
\end{algorithmic}
\end{algorithm}

Since we train each network separately and combine them only when choosing actions, it is no surprise that original DQN and Deep TAMER are written in this DQN-TAMER framework. DQN is equivalent when $\alpha_h = 0$ and Deep TAMER is equivalent when $\alpha_q = 0$.
Thus, DQN-TAMER can also be seen as a method for annealing DQN that is aided by including human feedback in the pure DQN algorithm.

In summary, we have four algorithms: (1) DQN, (2) DQN with naive reward shaping where feedback is added to environmental rewards, (3) Deep TAMER, and (4) our proposed DQN-TAMER algorithm.
In the following experiments, we compare these algorithms and show that DQN-TAMER outperforms the others in terms of learning speed and final agent performance.

\section{Experimental Settings \label{sec:experimental_setup}}

Two experiments were performed.
The first experiment aims to compare and analyze each algorithm in fair and wide settings.
We prepare programs as simulated human observers based on the four requirements described in Sec~\ref{sec: problem_formulation} (\textsc{binary, delay, stochasticity, unsustainability}).
Following Griffith et al.~\cite{NIPS2013_5187}, the simulated human gives feedback when certain conditions are satisfied for a given state and action.
The simulated approaches are appreciated because we can systematically test the performance of various algorithms with hyperparameters in a consistent setting.
Even with deep RL algorithms, whose performance can vary largely due to random seeds, we can fairly compare them by averaging the results from many runs.
We actually used a trimmed mean of results from 30 runs in all experiments for a reliable comparison.

We trained the agents in %
two game environments: \textit{Maze} and \textit{Taxi}.
As for a human observer in the simulated world, there are parameters which should be decided beforehand ($p_{\rm delay}$ for \textsc{delay}, $p_{\rm feedback}$ for \textsc{stochasticity} and $t_{\rm stop}$ for \textsc{unsustainability}).
As for the probability of the delay, $p_{\rm delay}$,
we assume it as $p_{delay}(0) = 0.3$, $p_{delay}(1) = 0.6$, $p_{delay}(2) = 0.1$, $p_{delay}(n) = 0 \,\,(n \geq 3)$. Because this true probability of the delay is unknown in reality, we assume the different one during training, which is given by  $p_{delay}(i) = 1/3$ for $i \in \{0,\cdots,2\}$ otherwise $p_{delay}(i) = 0$,
following Warnell et al.~\cite{DBLP:conf/aaai/WarnellWLS18}.
\if0
$\sum_{i=0}^{\infty} p_{\rm delay}(i) = 1$.
It is reasonable to suppose this distribution is similar to that reported by Hockley~\cite{hockley1984analysis}.
In detail, we derive $p_{\rm delay}(i)$ as follows:
\begin{eqnarray}
p_{\rm delay}(i) &=& P( i\times d \leq t < (i+1)\times d) \nonumber \\
&=& \int_{i\times d }^{(i+1)\times d} f(\tau) d\tau,
\end{eqnarray}
where $f(\tau)$ is the probability density function based on Hockley's result and $d$ is the time interval between two actions of the agent.
In the following experiments, we fix that $p_{delay}(0) = 0.3$, $p_{delay}(1) = 0.6$, $p_{delay}(2) = 0.1$, $p_{delay}(n) = 0 \,\,(n \geq 3)$.
\fi
%

Second, we built a real human-in-the-loop RL system to demonstrate the effectiveness of the proposed method in real applications.
The system uses a camera to perceive human faces and interpret them as human feedback using a deep neural network for facial expression recognition.
Even though such implicit feedback is actively inferred by the system, it learns maze navigation well.
We show the results from the demo in our complementary video.

\subsection{Maze}
\begin{figure}[t]
 \begin{center}
	\includegraphics[width=40mm]{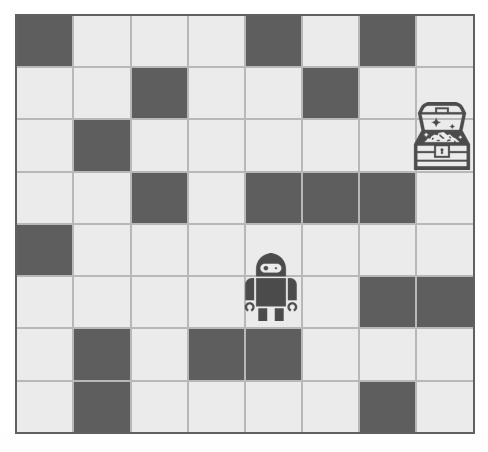}
 \end{center}
 \caption{Maze: an environment with walls (black squares), the agent, and the goal.}
 \label{fig:maze}
\end{figure}

Maze is a classical game where the agent must reach a predefined goal (Figure \ref{fig:maze}).
We compare the sample efficiency in each algorithm through experiment, i.e., we examine how fast learning converges.
We fixed the field size of a maze to 8 $\times$ 8 and the initial distance to the goal at 5.
Table \ref{tab:maze_setting} summarizes the environmental setting.

We simulate a human feedback as it gives a binary label whether the agent reduces the Manhattan distance to the goal.
If an agent moves closer to the goal, the human provides +1 positive feedback and -1 negative feedback otherwise.
We experimented with two different settings of observations $s_t$, which an agent can see from the environment.
In the first setting, an agent only knows its own absolute coordinates in a maze.
In the other setting, it observes the status of the surrounding areas (8 squares).
In the case shown in Figure \ref{fig:maze}, an agent observe either absolute coordinate ``(6, 5)'' or partial observation [``space'', ``space'', ``space'', ``space'', (``now'', ) ``space'
', ``wall'', ``wall'', ``space''] respectively in each setting.
Observation of only surrounding areas follows a partially observable Markov decision process (POMDP)~\cite{DBLP:journals/ai/KaelblingLC98}.
The POMDP framework is general enough to model a variety of real-world sequential decision processes, such as robot navigation problems, machine maintenance, and planning under uncertainty in general, but is also known it is difficult environment to train the agent.

\subsection{Taxi}

\begin{figure}[t]
 \begin{center}
	\includegraphics[width=40mm]{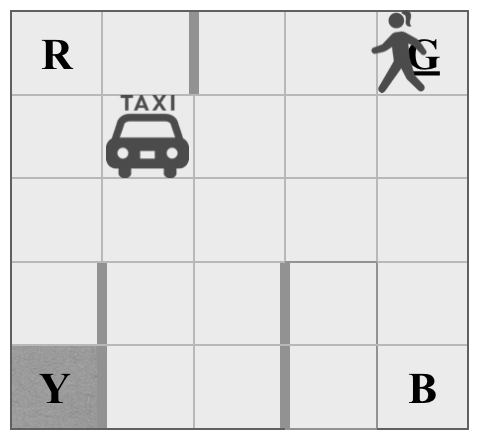}
 \end{center}
 \caption{Taxi: an environment with walls ( $|$ ; bold bars), the taxi agent, the passenger (at G), and the goal (Y).}
 \label{fig:taxi}
\end{figure}

Taxi is also a moving game in a two-dimensional space (Figure \ref{fig:taxi}), but it is more difficult due to its hierarchical goals~\cite{dietterich2000hierarchical}.
In Taxi, an agent must pick up a passenger that is waiting at a certain position and move him/her to a different position.
The position of the passenger and the final destination are randomly chosen from four candidate positions \{R, G, B, Y\}.

Thus, the optimal direction is different before and after picking up the passenger.
The agent must learn such a two-staged policy to solve this task.
We fix the field size of a maze to 5 $\times$ 5.
Table \ref{tab:taxi_setting} summarizes the environment settings.
The agent observes the current absolute coordinates and whether or not the passenger is currently in the taxi (agent). %
We simulate a human feedback as it gives a binary label whether it reduces the distance to a passenger or the destination according to the state of picking up.

\begin{table}[t]
  \centering
  \caption{Maze setting}
  \begin{tabular}{cc}\hline
   reward & every step -0.01, goal +1.0 \\ \hline
   field size & 8 \\ \hline
   initial distance & 5 \\ \hline
	max steps & 1000 \\ \hline
   action space & (north, east, south, west) \\ \hline
   human rule & Manhattan distance to the goal \\ \hline
  \end{tabular}
  \label{tab:maze_setting}
\end{table}

\begin{table}[t]
  \centering
  \caption{Taxi setting}
  \begin{tabular}{cc}\hline
   reward & 
   \begin{tabular}{c}
   	  every step -1, drop at right/wrong place +20/-10 \\
      pickup at the wrong place -10
   \end{tabular}     \\ \hline
   field size & 5 \\ \hline
   initial distance & random \\ \hline
	max steps & 1000 \\ \hline
   action space & (north, east, south, west, pickup, drop) \\ \hline
   human rule & 
   \begin{tabular}{c}
     Manhattan distance to passenger (before pick up) \\
     Manhattan distance to goal (after pick up)
   \end{tabular}     \\ \hline
  \end{tabular}
  \label{tab:taxi_setting}
\end{table}

\subsection{Car Robot Demonstration}

\begin{figure}[t]
 \begin{center}
	\includegraphics[width=80mm]{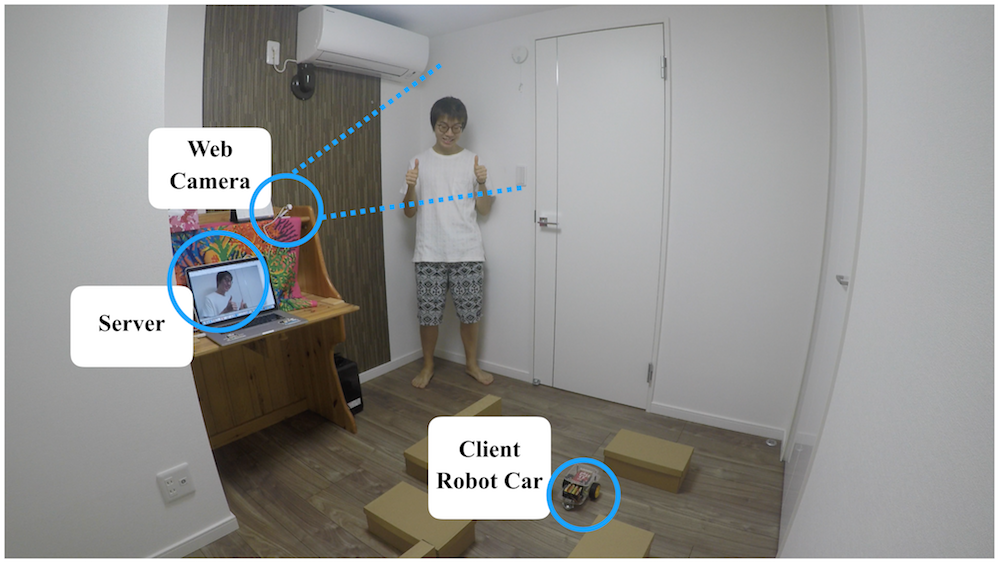}
 \end{center}
 \caption{Demonstration situation. We used a GoPiGo3 car robot and trained it to solve a maze using human facial expressions.}
 \label{fig:fig_demonstration}
\end{figure}

As a further demonstration, we built a demo system and trained a car agent with a real human observer.
We also introduce \textsc{natural reaction} in this demonstration as described in Sec.~\ref{sec: problem_formulation}, thus bringing the system closer to real applications.
Feedback is inferred by observing a person and is obtained through facial expression recognition.
We used MicroExpNet as a recognition model, which is a convolutional neural network-based (CNN) model~\cite{DBLP:journals/corr/abs-1711-07011}.
This model is obtained by distilling a larger CNN model,
which then quickly and accurately classifies facial expressions into 8 categories: {`neutral', `anger', `contempt', `disgust', `fear', `happy', `sadness', `surprise'}.
Even such an accurate model, of course, often fails to predict the correct expression.
The intriguing question we tackle here is whether an agent can learn well from suspicious feedback with errors.
Figure \ref{fig:fig_demonstration} shows how we set up the environment with a car robot solving a physical maze.
The agent interprets the facial expression `happy' as positive (+1) and other expressions (`anger', `contempt', `disgust', `fear', and `sad') as negative (-1).

\subsection{Parameter Settings}
We construct every Q-function and H-function as a feed-forward neural network with a hidden layer using tanh function of 100 dimensions.
Optimization is performed using RMSProp, where the initial learning rate is $10^{-3}$ both for the Q-function and the H-function.
The probability of taking random actions for exploration is initially set to $0.3$ and decayed by $0.001$ at every step until it reaches $0.1$.
We initialize $\alpha_h = \alpha_q = 1$ of the DQN-TAMER and decay $\alpha_h$ by $0.9999$ at every step.

\section{Results and Discussion \label{sec:results}}

In the following, we show the averaged results over totally 30 trials for each environment, where the results were obtained from three each with ten different sets of initial conditions.

\subsection{ \textsc{delay} and \textsc{stochasticity}} \label{subsec: delay_stochasticity}

\begin{figure}[t]
 \begin{minipage}{0.49\hsize}
  \begin{center}
   \includegraphics[width=45mm]{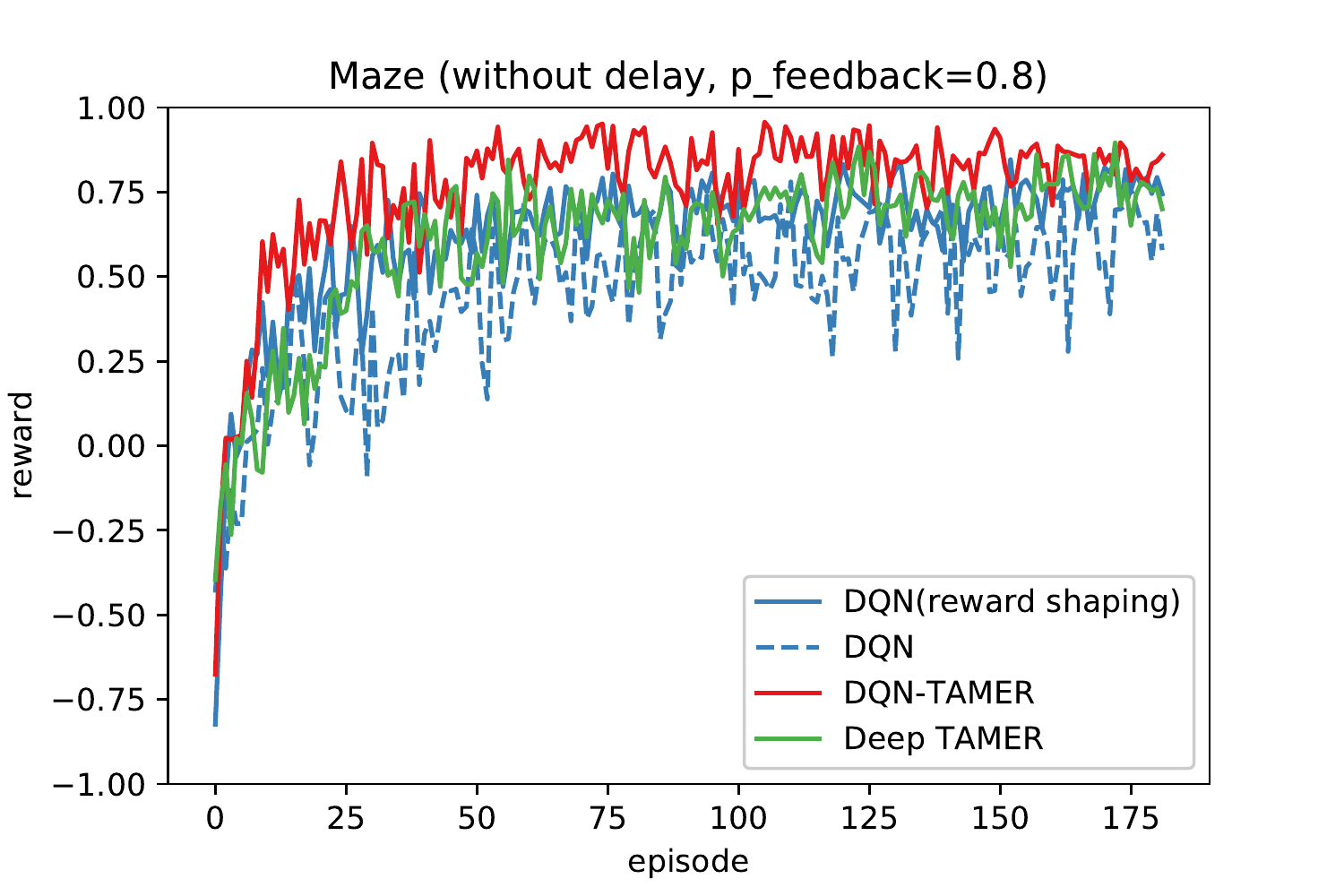}
  \end{center}
 \end{minipage}
 \begin{minipage}{0.49\hsize}
  \begin{center}
   \includegraphics[width=45mm]{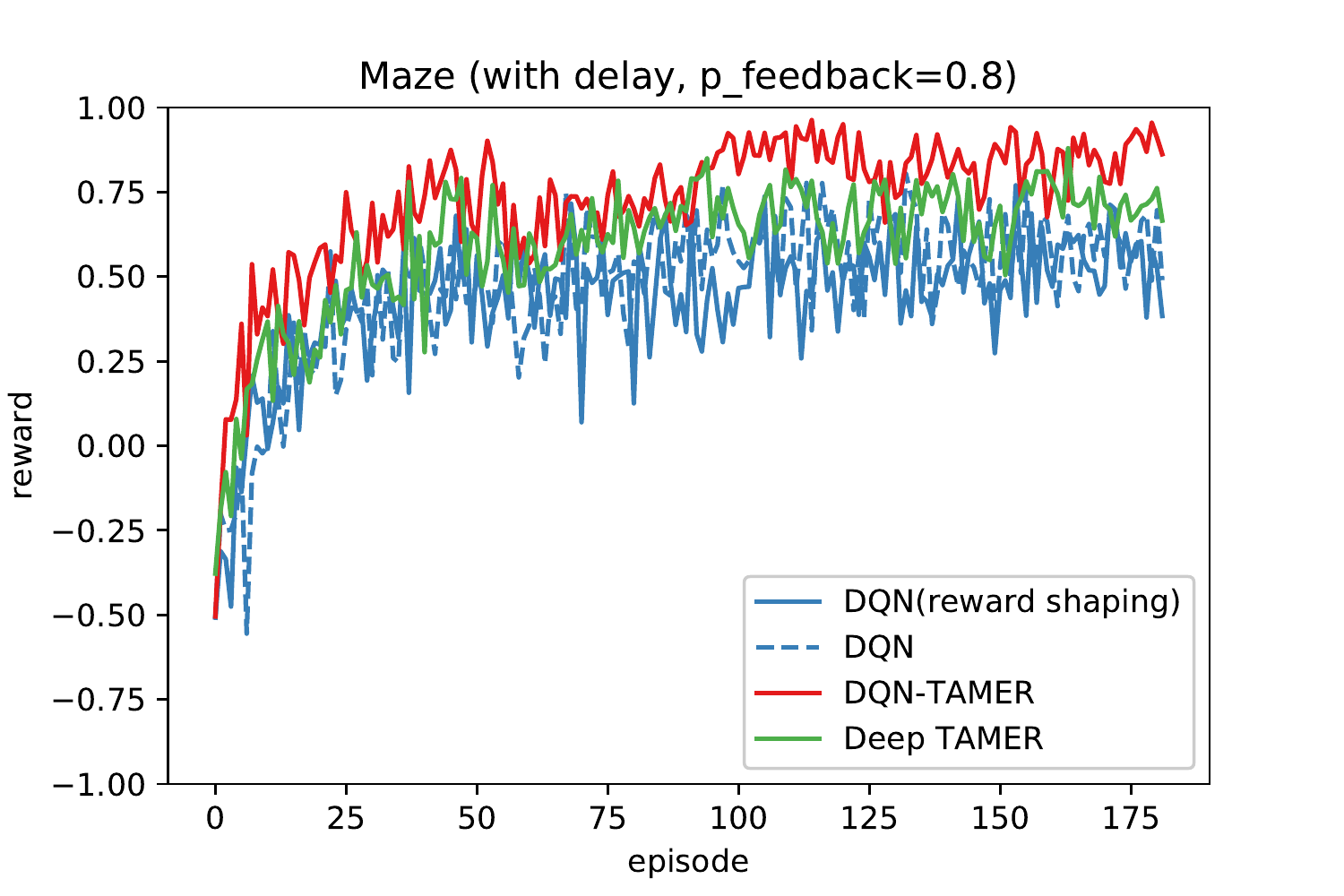}
  \end{center}
 \end{minipage} \\
 
 \begin{minipage}{0.49\hsize}
  \begin{center}
   \includegraphics[width=45mm]{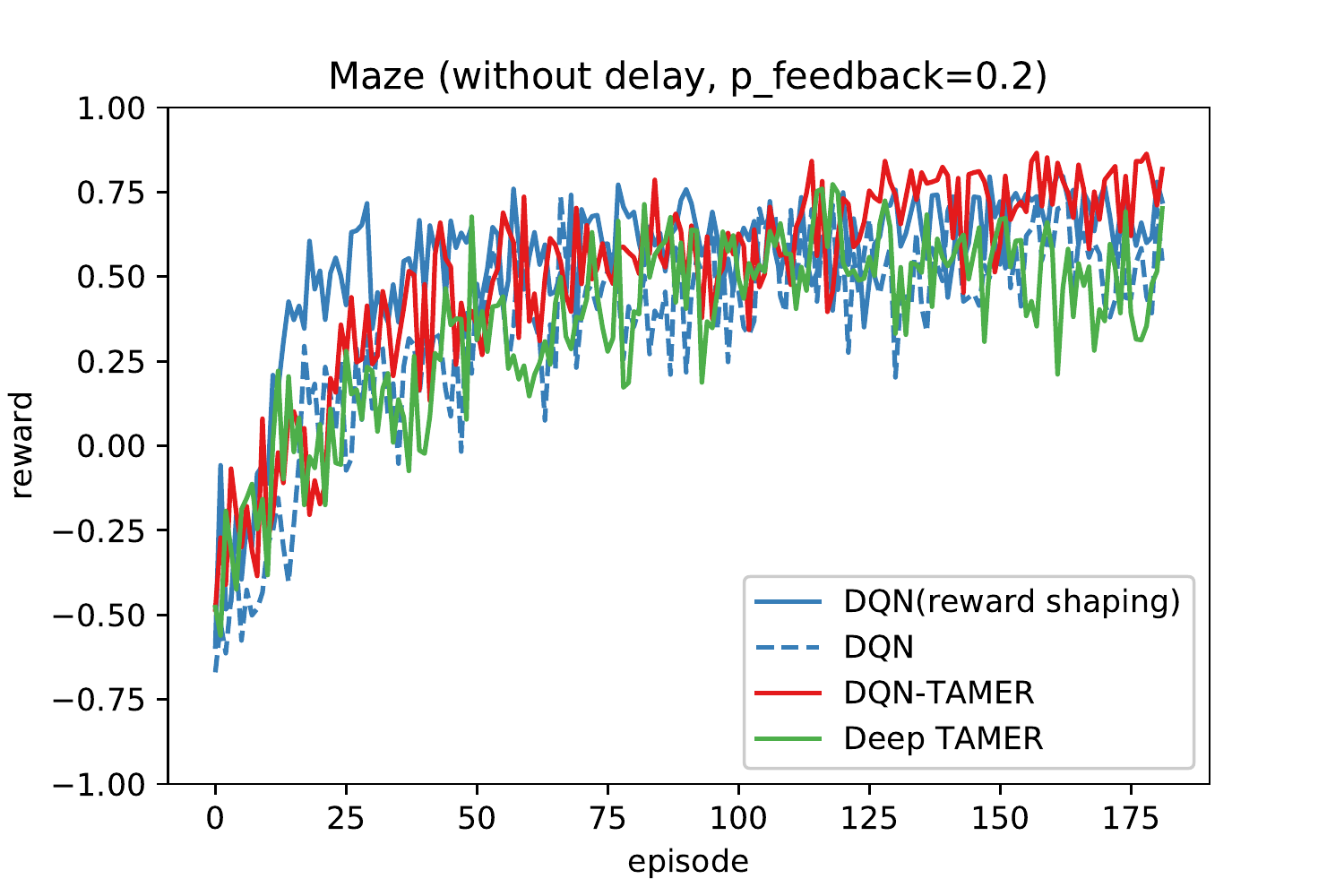}
  \end{center}
 \end{minipage}
 \begin{minipage}{0.49\hsize}
  \begin{center}
   \includegraphics[width=45mm]{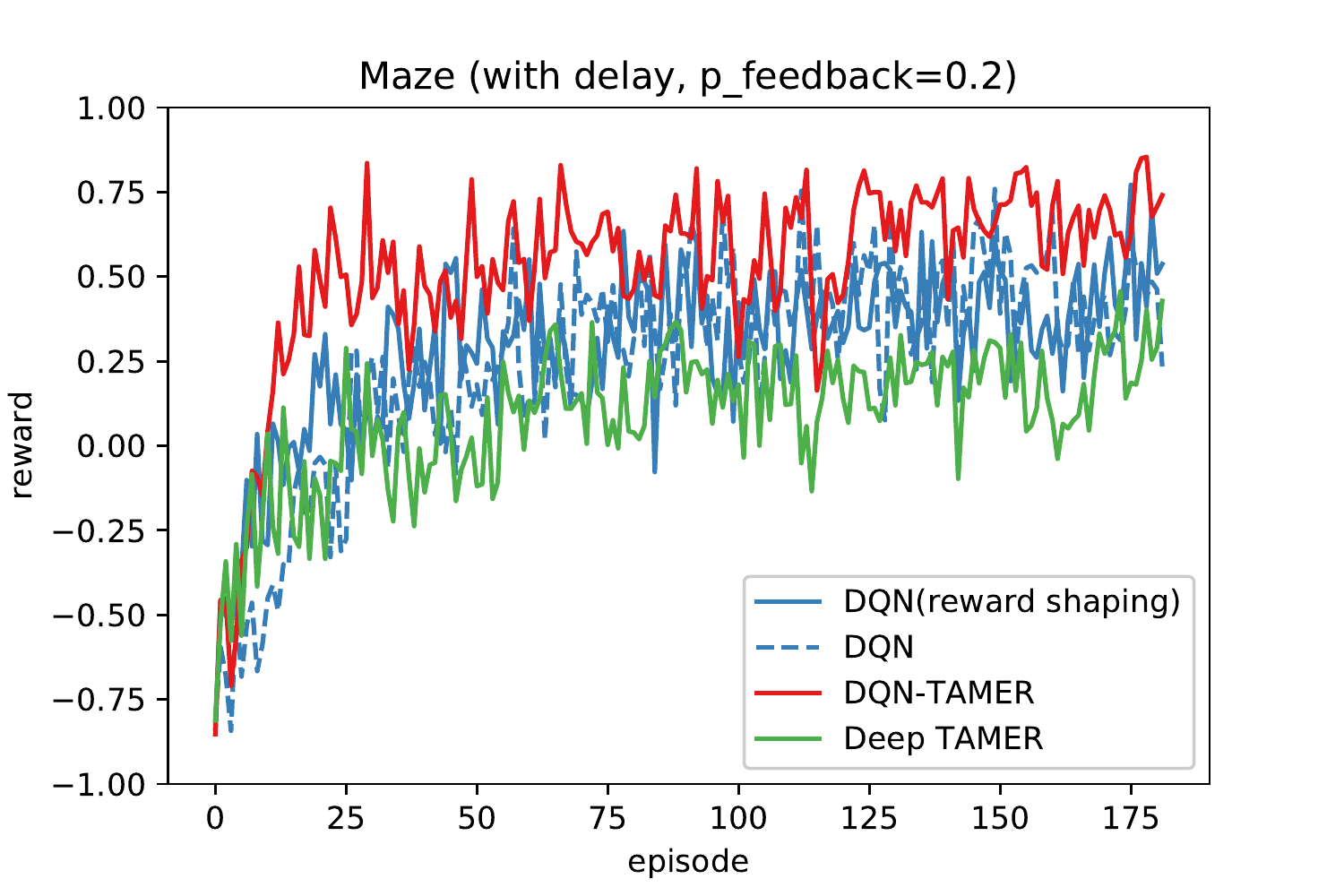}
  \end{center}
 \end{minipage}
  \caption{Maze results (upper: high frequency, lower: low frequency, left: without delay, and right: with delay).}
   \label{fig:maze_results}
\end{figure}

\begin{figure}[t]
 \begin{minipage}{0.49\hsize}
  \begin{center}
   \includegraphics[width=45mm]{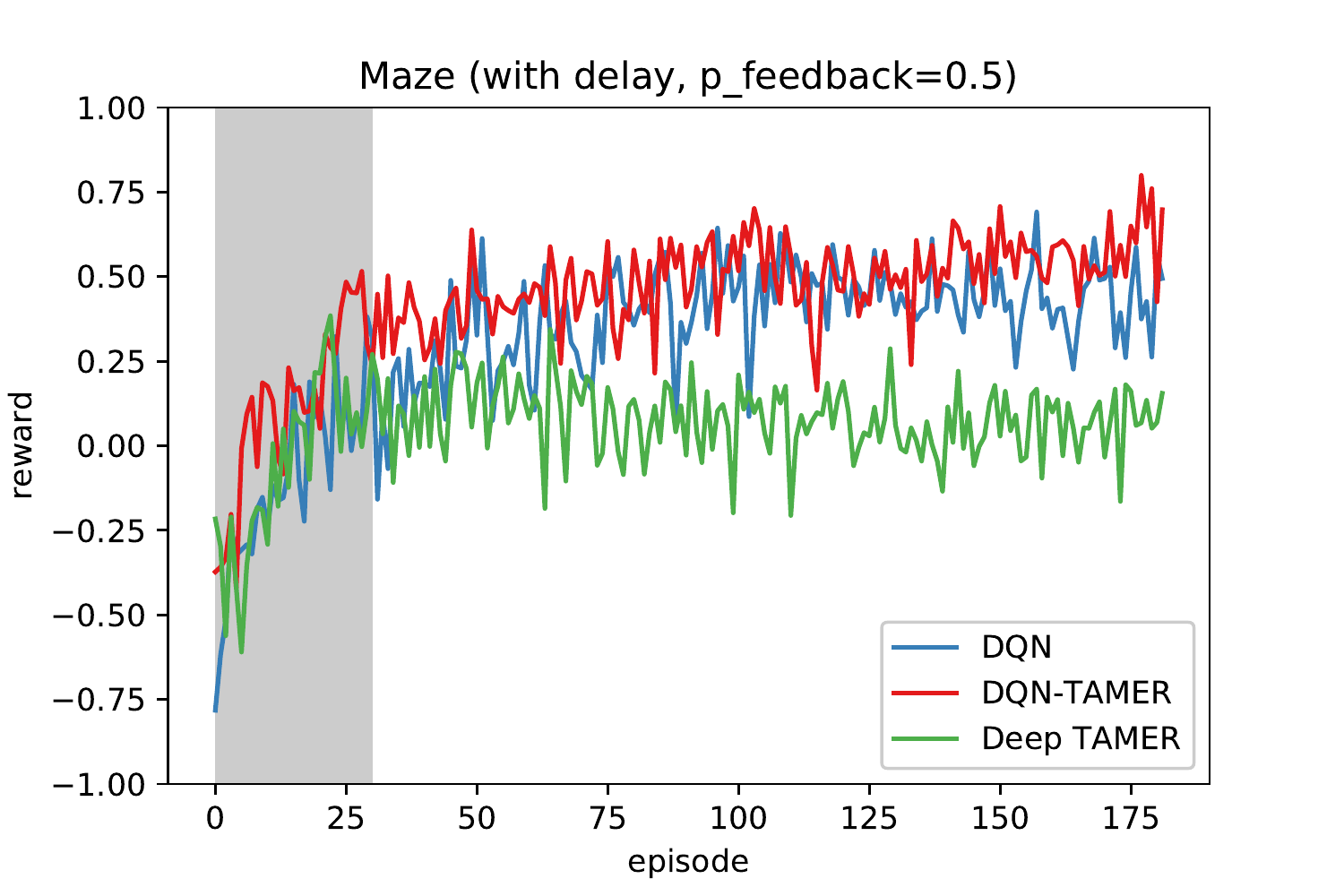}
  \end{center}
 \end{minipage}
 \begin{minipage}{0.49\hsize}
  \begin{center}
   \includegraphics[width=45mm]{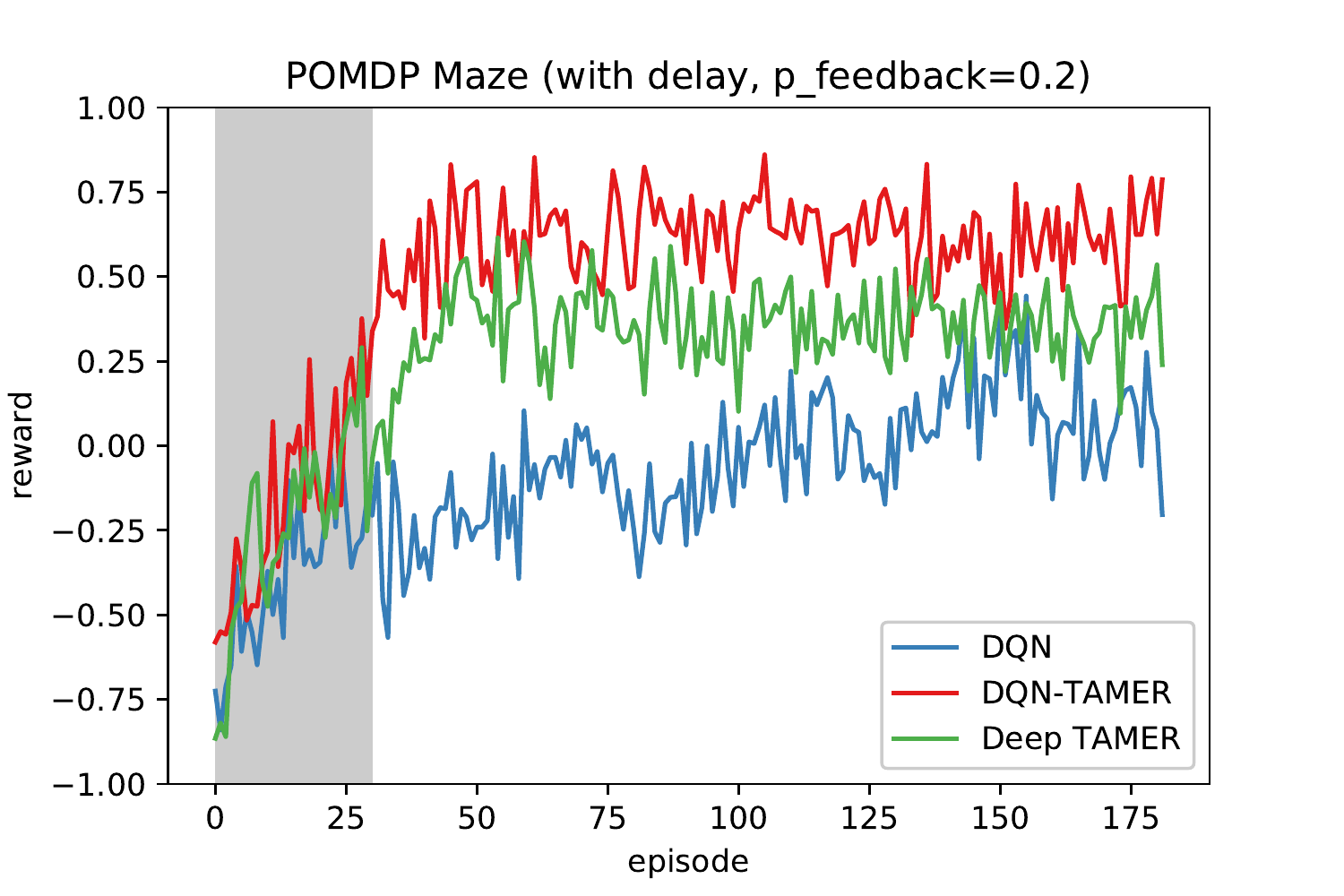}
  \end{center}
 \end{minipage}
  \caption{Maze with feedback stop. Feedback ends after 30 episodes. \\ left: MDP, right: POMDP}
   \label{fig:maze_high_freq_feedback}
\end{figure}

\begin{figure}[t]
 \begin{center}
	\includegraphics[width=50mm]{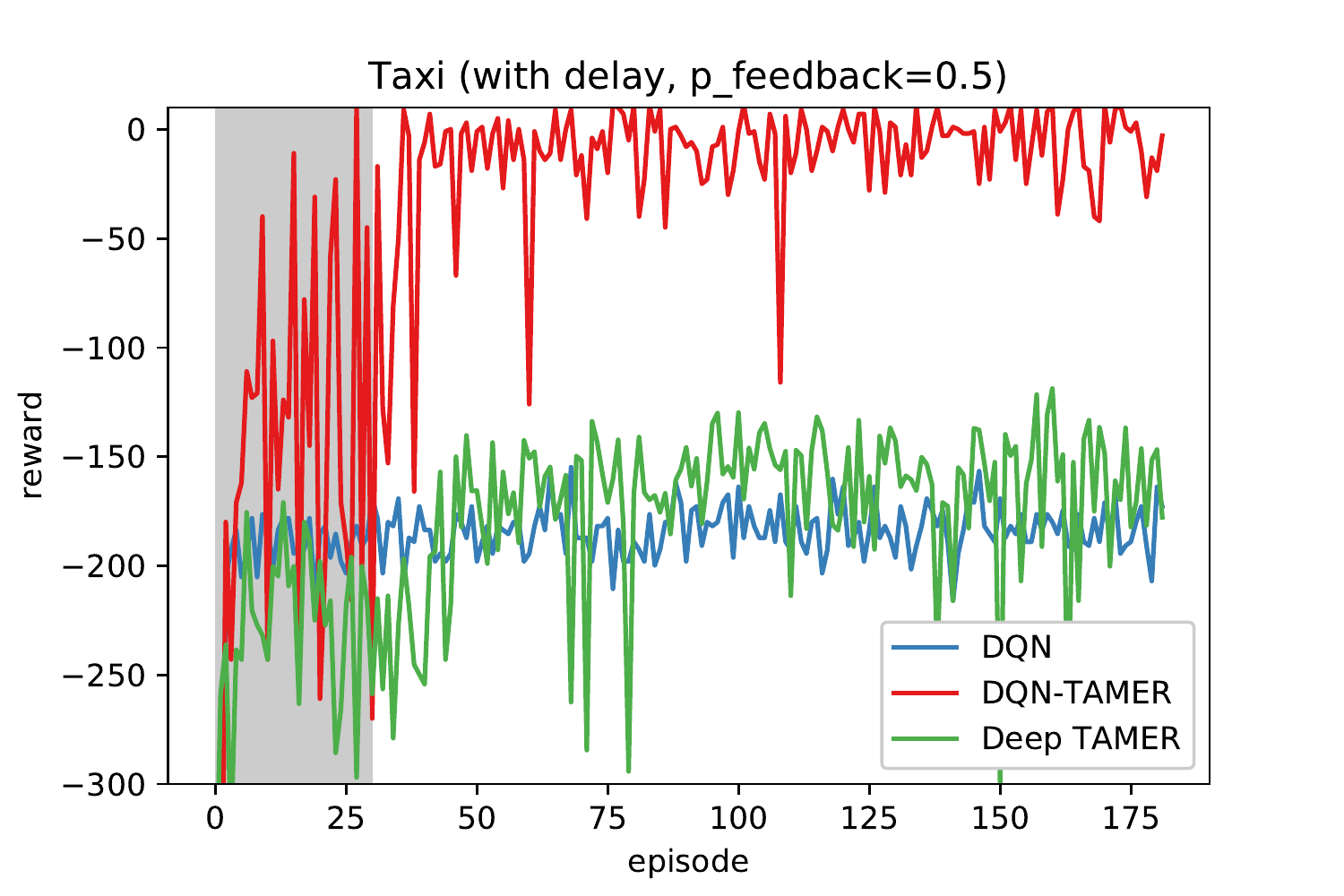}
 \end{center}
 \caption{Taxi with feedback stop. Feedback ends after 30 episodes.}
\end{figure}

To investigate the dependence on the delay of human feedback and the feedback occurrence probability,
we conducted experiments by varying the probability of feedback occurrence ($p_{\rm feedback}$) and the existence of delay.
Figure \ref{fig:maze_results} shows the four results of Maze
each of which corresponds the condition either the feedback frequency is high and low, and delay happens or does not.

As for \textsc{delay},
we can see that DQN with reward shaping outperforms DQN if there is no delay by comparing left and right panels of the figure.
However, the performance of one with reward shaping degrades and becomes comparable with pure DQN if delay is introduced.
This suggests that the human feedback
does not work well by naive reward shaping.

Comparing the upper and lower figures, one can see that a learning process with more frequent feedback is faster and reaches higher rewards for all algorithms.
Less frequent feedback degrades the performance of all algorithms.
Deep TAMER returned a particularly poor result.
Among those, DQN-TAMER is the most robust with unstable feedback since it uses a Q-function and an H-function. Therefore, it can also take advantage of rewards from the environment.

\subsection{\textsc{unsustainability}} \label{subsec: unsustainability}
We investigate the effect to learning when human feedback gets interrupted.
In any case, DQN-TAMER outperforms the other methods. It is inferred that Deep TAMER becomes stagnant after feedback stops because it depends only on human feedback.
In contrast, DQN-TAMER initially facilitates efficient exploration with human feedback and continues improving its policy with rewards from the environment.
The result is consistent with experimental results from Maze and Taxi.
Our proposed DQN-TAMER is very robust to various types of human feedback.

\subsection{\textsc{natural reaction}}
During the car robot demonstration, we found that the agent learned well from suspicious feedback with errors of the classifier efficiently.
The facial expression classifier misclassified human facial expressions (i.e., flipping plus and minus of reward) with around 15\%.
The result demonstrated that the DQN-TAMER was robust even though such opposite feedback occur stochastically.
We show the learning process in the complementary video.

\section{Conclusion}

This study tackles a series of challenges for introducing human-in-the-loop RL into real world robotics.
We discussed five key problems for human feedback in real applications: \textsc{binary}, \textsc{delay}, \textsc{stochasticity}, \textsc{unsustainability} and \textsc{natural reaction}.
The experiments results obtained from various settings show that the proposed DQN-TAMER model is robust against inconvenient feedback
and outperforms existing algorithms like DQN and Deep TAMER.
We also built a car robot system that exploits implicit rewards by reading human faces with a CNN based classifier.
Even with classifier errors, the agent of the system efficiently learned maze navigation.
These results encourage to utilize the human feedback in a real world scenario, which is difficult to handle due the instability and randomness of the delay, if we assume the randomness of the delay even when the probability function is different from the true one and combine the human feedback appropriately with the original reward given by the environment.

\newpage
\bibliographystyle{IEEEtran}
\bibliography{bibliography}

\begin{thebibliography}{10}
\providecommand{\url}[1]{#1}
\csname url@rmstyle\endcsname
\providecommand{\newblock}{\relax}
\providecommand{\bibinfo}[2]{#2}
\providecommand\BIBentrySTDinterwordspacing{\spaceskip=0pt\relax}
\providecommand\BIBentryALTinterwordstretchfactor{4}
\providecommand\BIBentryALTinterwordspacing{\spaceskip=\fontdimen2\font plus
\BIBentryALTinterwordstretchfactor\fontdimen3\font minus
  \fontdimen4\font\relax}
\providecommand\BIBforeignlanguage[2]{{%
\expandafter\ifx\csname l@#1\endcsname\relax
\typeout{** WARNING: IEEEtran.bst: No hyphenation pattern has been}%
\typeout{** loaded for the language `#1'. Using the pattern for}%
\typeout{** the default language instead.}%
\else
\language=\csname l@#1\endcsname
\fi
#2}}

\bibitem{DBLP:journals/ijrr/KoberBP13}
J.~Kober, \emph{et~al.}, ``Reinforcement learning in robotics: {A} survey,''
  \emph{I. J. Robotics Res.}, vol.~32, no.~11, pp. 1238--1274, 2013.

\bibitem{DBLP:journals/jmlr/LevineFDA16}
S.~Levine, \emph{et~al.}, ``End-to-end training of deep visuomotor policies,''
  \emph{Journal of Machine Learning Research}, vol.~17, pp. 39:1--39:40, 2016.

\bibitem{DBLP:conf/icra/GuHLL17}
S.~Gu, \emph{et~al.}, ``Deep reinforcement learning for robotic manipulation
  with asynchronous off-policy updates,'' in \emph{{IEEE} International
  Conference on Robotics and Automation, {ICRA}}, 2017, pp. 3389--3396.

\bibitem{DBLP:journals/corr/abs-1806-07857}
J.~A. Arjona{-}Medina, \emph{et~al.}, ``{RUDDER:} return decomposition for
  delayed rewards,'' \emph{CoRR}, vol. abs/1806.07857, 2018.

\bibitem{DBLP:journals/corr/abs-1709-10089}
A.~Nair, \emph{et~al.}, ``Overcoming exploration in reinforcement learning with
  demonstrations,'' \emph{CoRR}, vol. abs/1709.10089, 2017.

\bibitem{Thomaz2005RealTimeIR}
A.~L. Thomaz, \emph{et~al.}, ``Real-time interactive reinforcement learning for
  robots,'' in \emph{AAAI 2005 workshop on human comprehensible machine
  learning}, 2005.

\bibitem{DBLP:conf/ro-man/ThomazHB06}
------, ``Reinforcement learning with human teachers: Understanding how people
  want to teach robots,'' in \emph{The 15th {IEEE} International Symposium on
  Robot and Human Interactive Communication, {RO-MAN}}, 2006, pp. 352--357.

\bibitem{DBLP:conf/aihc/Broekens07}
J.~Broekens, ``Emotion and reinforcement: affective facial expressions
  facilitate robot learning,'' in \emph{Artifical intelligence for human
  computing}.\hskip 1em plus 0.5em minus 0.4em\relax Springer, 2007, pp.
  113--132.

\bibitem{4640845}
W.~B. Knox and P.~Stone, ``{TAMER}: Training an agent manually via evaluative
  reinforcement,'' in \emph{2008 7th IEEE International Conference on
  Development and Learning}, Aug 2008, pp. 292--297.

\bibitem{DBLP:conf/iberamia/Tenorio-GonzalezMP10}
A.~C. Tenorio-Gonzalez, \emph{et~al.}, ``Dynamic reward shaping: Training a
  robot by voice,'' in \emph{Proceedings of the 12th Ibero-American Conference
  on Advances in Artificial Intelligence}, 2010, pp. 483--492.

\bibitem{5975338}
P.~M. Pilarski, \emph{et~al.}, ``Online human training of a myoelectric
  prosthesis controller via actor-critic reinforcement learning,'' in
  \emph{IEEE International Conference on Rehabilitation Robotics}, 2011, pp.
  1--7.

\bibitem{NIPS2013_5187}
S.~Griffith, \emph{et~al.}, ``Policy shaping: Integrating human feedback with
  reinforcement learning,'' in \emph{Advances in Neural Information Processing
  Systems 26}, 2013, pp. 2625--2633.

\bibitem{DBLP:conf/icml/MacGlashanHLPWR17}
J.~MacGlashan, \emph{et~al.}, ``Interactive learning from policy-dependent
  human feedback,'' in \emph{Proceedings of the 34th International Conference
  on Machine Learning}, 2017, pp. 2285--2294.

\bibitem{Arumugam2018DeepRL}
D.~Arumugam, \emph{et~al.}, ``Deep reinforcement learning from policy-dependent
  human feedback,'' 2018.

\bibitem{DBLP:conf/aaai/WarnellWLS18}
G.~Warnell, \emph{et~al.}, ``Deep {TAMER}: Interactive agent shaping in
  high-dimensional state spaces,'' in \emph{Proceedings of the Thirty-Second
  {AAAI} Conference on Artificial Intelligence}, 2018.

\bibitem{PRESTON20001}
C.~C. Preston and A.~M. Colman, ``Optimal number of response categories in
  rating scales: reliability, validity, discriminating power, and respondent
  preferences,'' \emph{Acta Psychologica}, vol. 104, no.~1, pp. 1 -- 15, 2000.

\bibitem{DBLP:conf/nips/ChristianoLBMLA17}
P.~F. Christiano, \emph{et~al.}, ``Deep reinforcement learning from human
  preferences,'' in \emph{Advances in Neural Information Processing Systems
  30}, 2017, pp. 4302--4310.

\bibitem{hockley1984analysis}
W.~E. Hockley, ``Analysis of response time distributions in the study of
  cognitive processes.'' \emph{Journal of Experimental Psychology: Learning,
  Memory, and Cognition}, vol.~10, no.~4, p. 598, 1984.

\bibitem{DBLP:conf/nips/IsbellSKSS01}
C.~L. Isbell~Jr and C.~R. Shelton, ``Cobot: A social reinforcement learning
  agent,'' in \emph{Advances in Neural Information Processing Systems}, 2002,
  pp. 1393--1400.

\bibitem{DBLP:conf/agents/IsbellSKSS01}
C.~Isbell, \emph{et~al.}, ``A social reinforcement learning agent,'' in
  \emph{Proceedings of the fifth international conference on Autonomous
  agents}.\hskip 1em plus 0.5em minus 0.4em\relax ACM, 2001, pp. 377--384.

\bibitem{Thrun1995LifelongRL}
S.~Thrun and T.~M. Mitchell, ``Lifelong robot learning,'' \emph{Robotics and
  Autonomous Systems}, vol.~15, pp. 25--46, 1995.

\bibitem{finn2016generalizing}
C.~Finn, \emph{et~al.}, ``Generalizing skills with semi-supervised
  reinforcement learning,'' in \emph{Proceedings of ICLR}, 2016.

\bibitem{Watkins1992}
C.~J. C.~H. Watkins and P.~Dayan, ``Q-learning,'' \emph{Machine Learning},
  vol.~8, no.~3, pp. 279--292, May 1992.

\bibitem{Sutton1988}
R.~S. Sutton, ``Learning to predict by the methods of temporal differences,''
  \emph{Machine Learning}, vol.~3, no.~1, pp. 9--44, Aug 1988.

\bibitem{DBLP:books/lib/RussellN03}
S.~J. Russell and P.~Norvig, \emph{Artificial intelligence - a modern approach,
  2nd Edition}, ser. Prentice Hall series in artificial intelligence, 2003.

\bibitem{Lin1992}
L.-J. Lin, ``Self-improving reactive agents based on reinforcement learning,
  planning and teaching,'' \emph{Machine Learning}, vol.~8, no.~3, pp.
  293--321, May 1992.

\bibitem{DBLP:journals/corr/MnihKSGAWR13}
V.~Mnih, \emph{et~al.}, ``Playing atari with deep reinforcement learning,''
  \emph{CoRR}, vol. abs/1312.5602, 2013.

\bibitem{DBLP:conf/icml/NgHR99}
A.~Y. Ng, \emph{et~al.}, ``Policy invariance under reward transformations:
  Theory and application to reward shaping,'' in \emph{Proceedings of the
  Sixteenth International Conference on Machine Learning}, 1999, pp. 278--287.

\bibitem{DBLP:conf/iui/KnoxS13}
W.~B. Knox and P.~Stone, ``Learning non-myopically from human-generated
  reward,'' in \emph{18th International Conference on Intelligent User
  Interfaces}, 2013, pp. 191--202.

\bibitem{DBLP:journals/ai/KaelblingLC98}
L.~P. Kaelbling, \emph{et~al.}, ``Planning and acting in partially observable
  stochastic domains,'' \emph{Artif. Intell.}, vol. 101, no. 1-2, pp. 99--134,
  1998.

\bibitem{dietterich2000hierarchical}
T.~G. Dietterich, ``Hierarchical reinforcement learning with the maxq value
  function decomposition,'' \emph{Journal of Artificial Intelligence Research},
  vol.~13, pp. 227--303, 2000.

\bibitem{DBLP:journals/corr/abs-1711-07011}
I.~{\c{C}}ugu, \emph{et~al.}, ``Microexpnet: An extremely small and fast model
  for expression recognition from frontal face images,'' \emph{arXiv}, vol.
  1711.07011, pp. 1--9, 2017.

\end{thebibliography}

\end{document}